\documentclass[preprint]{aastex}
\newcommand{\simle}{\mbox{$\stackrel{<}{_{\sim}}$}}
\newcommand{\simge}{\mbox{$\stackrel{>}{_{\sim}}$}}

\newcommand\etal{ {\em et~al.\/}\thinspace}

\newcommand\msun{\hbox{\,M$_\odot$}}

\slugcomment{To appear 2002Feb10 Astrophysical Journal}

\shorttitle{Radio Pinwheels}
\shortauthors{Monnier et al.}

\begin{document}

\title{Radio Properties of Pinwheel Nebulae}


\author{J. D. Monnier\altaffilmark{1}, L. J. Greenhill\altaffilmark{1}, 
P. G. Tuthill\altaffilmark{2} and W. C. Danchi\altaffilmark{3}}

\altaffiltext{1}{Harvard-Smithsonian Center for Astrophysics, MS\#42,
60 Garden Street, Cambridge, MA, 02138}
\altaffiltext{2}{School of Physics, University of Sydney, NSW 2006, Australia}
\altaffiltext{3}{NASA Goddard Space Flight Center,
Infrared Astrophysics, Code 685, Greenbelt, MD 20771}
\email{jmonnier@cfa.harvard.edu, lincoln@cfa.harvard.edu, gekko@physics.usyd.edu.au, wcd@iri1.gsfc.nasa.gov}


\begin{abstract}

A small number of dusty Wolf-Rayet stars have been resolved into
pinwheel nebulae, defined by their ``rotating'' spiral dust shells
observed in the infrared.  This morphology is naturally explained by
dust formation associated with colliding winds in a binary system.  In
order to confirm and further explore this hypothesis, we have observed
the known pinwheel nebulae (WR~104 and WR~98a) as well as the
suspected binary WR~112 at multiple radio wavelengths with the Very
Large Array to search for non-thermal radio emission from colliding
winds.  The spectrum of each target is nearly flat between 5 and 22~GHz,
consistent with the presence of non-thermal emission that is
reduced at low frequencies by free-free absorption.  This emission
must lie outside the radio ``photosphere,'' leading us to estimate a
lower limit to the physical size of the non-thermal emitting region
that is larger than expected from current theory.  Based on a radio
and infrared comparison to WR~104 \& 98a, we conclude that WR~112 is a
likely candidate pinwheel nebula, but its temporal variability
indicates an eccentric binary orbit or a pinwheel viewed nearly
edge-on.  A sensitive radio survey of IR-bright WRs would
stringently test the hypothesis that colliding winds lie at the heart
of {\em all} dusty WR systems.  We also discuss the effects of dust
obscuration in the ultra-violet and how radio-determined mass-loss rates
of pinwheel nebulae (and dusty WR stars in general) 
may be underestimated due to shadowing effects.
\end{abstract}



%\tableofcontents

\section{Introduction}
The vast majority of late-WC (carbon-rich) Wolf-Rayet (WR) stars are
surrounded by dust shells \citep{williams87} that absorb stellar flux
and re-emit the energy in the infrared (IR).  Recently, the geometry
of the IR-brightest dust shells have been resolved into spiral plumes
that rotate with a $\sim$1~yr period: the pinwheel nebulae
\citep{tuthill99,monnier99}.  This morphology has been explained as a
consequence of colliding winds with unequal momenta, where dust forms
at the interface or in the wake 
of the winds and is subsequently carried out in the
flow \citep{usov91}.  
Because of orbital motion, the direction of the dust flow
rotates, generating the observed spiral pattern although the dust motion
itself is purely radial. By analyzing the time-dependent morphology,
important orbital and wind parameters, such as the period, inclination
angle, eccentricity, and even the distance (when combined with
observed terminal wind speeds), can be determined with 
high precision \citep{mythesis,winds2001b}.

These observations have led to a more unified picture of the dusty
Wolf-Rayet stars in terms of interacting wind (binary) systems.
\citet{wh92} have categorized dusty WC's as either ``variable'' or
``persistent'' dust-producers, based on the variability of IR flux.
WR~140 and other variable sources consist of
eccentric systems with $\ga$10~year orbits catalyzing dust formation
only near periastron \citep{moffat87,williams87b,williams90}, 
while WR~104 and
WR~98a have more circular orbits \citep[possibly circularized at an
earlier epoch;][]{monnier99} with periods $\sim$1~year,
allowing continuous (``persistent'') dust production.  An important
unanswered question is whether all dusty WR sources lie somewhere
along a continuum of binary orbits \citep{wh92}, or whether some of them could
be single stars
\citep{zubko98}.

Another potential observational signature of  binaries in dusty WR systems
is non-thermal
emission from colliding winds.  However, one must first establish that 
non-thermal radio emission from a WR star requires a
massive companion \citep{vdHucht92}. 
Recently, \citet{dw2000} re-examined just this question,
and they found that known WR binaries with periods longer
than 1\,year do consistently show evidence for non-thermal radio
emission, supporting the theory that colliding winds are responsible
\citep{eu93,jardine96}.  Further, \citet{dw2000} show that most 
binaries with periods less than 1\,year
have radio spectra similar to that expected for pure thermal
wind sources (spectral index $\alpha^T \sim 0.65$), 
the intrinsic non-thermal emission presumably having been mostly
absorbed by the ionized wind.

Having periods ($\sim$1~year) lying between the short and long period
systems studied previously, 
it is unclear whether pinwheel nebulae should
have detectable non-thermal radio emission.
We have observed 
the confirmed pinwheel nebulae around WR~104 and WR~98a and the
enigmatic candidate pinwheel WR~112 with the Very Large
Array (VLA) of the National Radio Astronomy 
Observatory\footnote{The National Radio Astronomy Observatory is a facility
of the National Science Foundation operated under cooperative
agreement by Associated Universities, Inc.} in order to characterize
their radio properties.  The goals of this study were to confirm the
colliding wind origin of the pinwheel nebulae by definitive detection
of non-thermal radio emission, to determine if there exists a
distinctive radio signature that can be used as a diagnostic for binarity
in systems whose infrared structure is too small or too dim to resolve
\citep{winds2001a}, and to elucidate the true nature of the
WR~112 system whose IR and radio properties have been difficult to
understand \citep{rr99,monnier2001a}.

\section{Observations and Results}
\label{section:observations}
We have used the Very Large Array (VLA) to measure the broadband
spectra of three dust-enshrouded Wolf-Rayet systems, WR~104, WR~98a,
and WR~112, at 1.43, 4.86, 8.46, 14.9, \& 22.5\,GHz.  
Before beginning our study, we
cross-referenced the persistent and variable WR dust emitters from
\citet{wh92} (which includes our target sample) with the southern
hemisphere sample of \citet{leitherer97} and \citet{chapman99}.  
Of the 12 sources in common, 
only two have been detected in the radio (WR~65 and WR~112).  
In order to have good chance to detect our program
sources at most of the target wavelengths,
our program
was designed to improve upon the sensitivity limits of these previous
surveys by factors of 3 to 10, with a goal of $\la$100\,$\mu$Jy
\footnote{1~Jy = 10$^{-26}$~W~m$^{-2}$~Hz$^{-1}$}
point-source sensitivity.  The data reported here were collected
during two separate observing runs, in 1999 September and 2000
February, when the VLA was in AnB and BnC configurations respectively; 
3C286 was used for primary amplitude calibration.
Poor observing conditions during the September epoch required some
observations to be repeated in February.  A preliminary report of
our results appeared in \citet{winds2001a}.

Since we were observing at five frequency bands, a number of
secondary phase calibrators were used, and a complete journal of our
calibrator observations appears in Table\,\ref{table:caljournal},
including the adopted VLA calibrator positions and our new flux density
measurements.  In general we followed the guidelines of the VLA
calibrator manual for amplitude calibration;
%(see \url{
%http://info.aoc.nrao.edu/~gtaylor/calman/boot.html}).
however for WR 104 and
WR 98a, we used the compact component of Sgr A$\ast$ as a
high-frequency calibrator by considering only (u,v) components beyond
50 k$\lambda$ (100 k$\lambda$) at 8.46 \& 14.9\,GHz (22.5\,GHz).  Our 8.46\,GHz
flux estimates for Sgr A$\ast$ are similar to contemporaneous
measurements by G. C. Bower (personal communication, 2000), but our fluxes at
higher frequencies
are $\sim$15\% higher.  This discrepancy may be due to
slightly different cutoffs in (u,v) coverage 
employed during calibration or variability of Sgr~A$\ast$
itself \citep{zhao2001}, but is also
nearly consistent with the expected level of systematic errors in the
amplitude calibration; our final conclusions do
not hinge upon this possible source of miscalibration.

\begin{table}[t]
\begin{center}
\scriptsize
\caption{Journal of calibrator observations \label{table:caljournal} }
%\begin{tabular}{|l|ccc|lccc|l|}
\begin{tabular}{lccccccccl}
\tableline\tableline

	 & \multicolumn{2}{c}{Position (J2000)} & Date   &
 \multicolumn{5}{c}{Measured Flux Density\tablenotemark{a,b} \quad (Jy) at} & Target\\
Source   &    RA  & Dec                         & (U.T.) & 1.43 GHz & 4.86 GHz & 8.46 GHz & 14.9 GHz  & 22.5 GHz & Calibrated \\
\tableline
1733-130    & 17 33 02.7058 & -13 04 49.548 & 
 1999 Sep 28 & ... & 4.416 & ... & 3.765 & 3.693 & WR~112\\
&&&
 1999 Sep 29 & ... & ... & ... & 3.833 & ... &  \\
&&&
 2000 Feb 15 & ... & ... & ... & 3.732 & 3.312 &  \\
\tableline
Sgr A$\ast$\tablenotemark{c} & 17 45 40.0409 & -29 00 28.118 &  
 1999 Sep 27 & ... & ... & 0.690 & 0.967 & 1.062 & WR~104, WR~98a\\
&&&
 1999 Sep 28 & ... & ... & 0.744 & 0.869 & 1.014 & \\
&&&
 2000 Feb 24 & ... & ... & 0.962 & 0.928 & ... & \\
&&&
 2000 Feb 25 & ... & ... & ... & ... & 1.003 & \\
\tableline
1751-253    & 17 51 51.2628 & -25 24 00.063 &
 1999 Sep 28 & 1.236 & ... & ... & ... & ... & WR~104, WR~98\\
&&&
 1999 Sep 29 & 1.244 & ... & ... & ... & ... & WR~112\\
&&&
 2000 Feb 15 & 1.229 & ... & ... & ... & ... & WR~112\\
\tableline
1820-254    & 18 20 57.8487 & -25 28 12.584 &
 1999 Sep 28 & ... & 1.094 & ... & ... & ... & WR~98a\\
\tableline
1832-105\tablenotemark{d}    & 18 32 20.8360 & -10 35 11.200 &
 1999 Sep 27 & ... & ... & 1.328 & ... & ... & WR~112\\
&&&
 1999 Sep 28 & ... & 1.174 & ... & ... & ... & \\
&&&
 2000 Feb 15 & ... & 1.321 & 1.359 & ... & ... & \\
\tableline
\end{tabular}
\tablenotetext{a}{Flux density scale was defined by the 
adopted strength of
3C286 using NRAO values of
14.75, 7.486, 5.181, 3.428, and 2.498\,Jy at
1.43, 4.86, 8.46, 14.9, \& 22.5\,GHz respectively.
}
\tablenotetext{b}{Statistical uncertainties 
were $\simle$1\%.
From the apparent scatter,
we estimate 
systematic calibration uncertainties were generally better
than 5\%, but occasionally as bad as 10\%.}
\tablenotetext{c}{Sgr A$\ast$ J2000 positions taken from \citet{reid99} measured
on 1996.25 not correcting for apparent proper motion; this should be 
accurate to $\sim$20~mas for the epoch of measurements reported here.}
\tablenotetext{d}{Positional accuracy $<$150~mas. 
All others here have $<$10~mas error.}
\end{center}
\end{table}

Table\,\ref{table:wrjournal} contains the calibrated fluxes and 
source positions for the target sample, including the first radio detections
of WR~104 and WR~98a.  The resolution of the VLA in A-array is sufficient to
resolve the thermal emission from some WR stars \citep[e.g.,][]{cr99}, 
and the data were inspected for evidence of structure.
Gaussian fits to the deconvolved images did not reveal
any significant deviations from point sources.

We note that the 8.46\,GHz measurement of WR~112 in 1999 September was corrupted
by poor atmospheric phase stability, and the bispectrum was vector-averaged 
in order to extract an estimate
of the point-source flux
\citep[for further discussion, see][]{cornwell87}.
This method is accurate only if no other sources are present in the field and
if the target is unresolved, and the validity of these
assumptions was checked using
subsequent observations in 2000 February.

\begin{table}[thb] 
\begin{center}
\scriptsize
\caption{Journal of Wolf-Rayet observations \label{table:wrjournal} }
\begin{tabular}{lcccccl}
\tableline\tableline
&        & Observing    & & & &  \\
&Date    & Frequency   & Measured Flux\tablenotemark{a,b,c} & 
\multicolumn{2}{c}{Position (J2000)\tablenotemark{d}} &  \\
Target Star & (U.T.)  & Band (GHz)  & Density (mJy) & RA & Dec & Comments\tablenotemark{e} \\

\tableline
WR 104 & 1999 Sep 28& L (1.425) & $<$0.30 & - & - & \\
       & 2000 Feb 24& X (8.460) & 0.87$\pm$0.06 & 
18 02 04.128$\pm$0.006 & -23 37 42.14$\pm$0.05 & Unresolved\\
       & 2000 Feb 24& U (14.94) & 1.02$\pm$0.12 & 
18 02 04.110$\pm$0.005 & -23 37 42.40$\pm$0.07 & Unresolved\\
       & 1999 Sep 28& K (22.46) & 0.90$\pm$0.15 & 
18 02 04.127$\pm$0.001 & -23 37 42.18$\pm$0.01 & Unresolved \\
       & 2000 Feb 25& K (22.46) & 0.97$\pm$0.12 & 
18 02 04.126$\pm$0.005 & -23 37 42.25$\pm$0.08 & Unresolved \\

\tableline
WR 98A & 1999 Sep 28& L (1.425) & $<$0.36 & - & - & \\
       & 1999 Sep 28& C (4.860) & 0.37$\pm$0.07 &
17 41 13.044 & -30 32 30.25 & Weak Detection\\
       & 1999 Sep 28& X (8.460) & 0.55$\pm$0.15 & 
17 41 13.044$\pm$0.005 & -30 32 30.35$\pm$0.10 & Weak Detection \\
       & 2000 Feb 24& X (8.460) & 0.62$\pm$0.05 &
17 41 13.057$\pm$0.006 & -30 32 30.39$\pm$0.06 & Unresolved\\
       & 2000 Feb 24& U (14.94) & 0.64$\pm$0.11 &
17 41 13.047$\pm$0.008 & -30 32 30.23$\pm$0.18 & \\
       & 2000 Feb 25& K (22.46) & 0.57$\pm$0.10 &
17 41 13.054$\pm$0.008 & -30 32 30.38$\pm$0.10 & \\

\tableline
WR 112 & 1999 Sep 29& L (1.425) & 2.71$\pm$0.17 &
18 16 33.484$\pm$0.002 & -18 58 42.79$\pm$0.03 & Unresolved \\
       & 2000 Feb 15& L (1.425) & 2.3$\pm$0.30 & 
18 16 33.484$\pm$0.006 & -18 58 42.36$\pm$0.13 & Unresolved \\
       & 1999 Sep 28& C (4.860) & 4.12$\pm$0.10 & 
18 16 33.488$\pm$0.030 & -18 58 42.47$\pm$0.30 & Unresolved\tablenotemark{f} \\
       & 2000 Feb 15& C (4.860) & 3.75$\pm$0.08 &
18 16 33.490$\pm$0.001 & -18 58 42.33$\pm$0.02 & Unresolved \\
       & 1999 Sep 27& X (8.460) & 4.4$\pm$0.3 &
- & - & Bispectrum\tablenotemark{g} \\
       & 2000 Feb 15& X (8.460) & 4.07$\pm$0.06 & 
18 16 33.4879$\pm$0.0003 & -18 58 42.289$\pm$0.006 & Unresolved \\
       & 1999 Sep 29& U (14.94) & 4.2$\pm$0.3 &
18 16 33.490$\pm$0.001 & -18 58 42.50$\pm$0.01 & Unresolved\\
       & 2000 Feb 15& U (14.94) & 4.39$\pm$0.17 & 
18:16 33.4902$\pm$0.0004 & -18 58 42.347$\pm$0.010 & Unresolved \\
       & 1999 Sep 28& K (22.46) & 4.05$\pm$0.25 &
18 16 33.4912$\pm$0.0001 & -18 58 42.353$\pm$0.002 & Unresolved \\
       & 2000 Feb 15& K (22.46) & 3.97$\pm$0.12 & 
18 16 33.4901$\pm$0.0003 & -18 58 42.374$\pm$0.007 & Unresolved \\
\tableline
\end{tabular}
\tablenotetext{a}{Flux density determined using AIPS tasks JMFIT and MAXFIT.}
\tablenotetext{b}{When not detected, we report 3-$\sigma$ upper limits.}
\tablenotetext{c}{Flux uncertainties here include only measurement error, 
not systematic calibration uncertainties which are generally $\sim$5\%.}
\tablenotetext{d}{Position estimates and errors derived from fits to a 
single Gaussian.}
\tablenotetext{e}{``Unresolved'': Gaussian fit to image consistent with
point source response (FWHM less than one-third of beam).  Blank: too 
little flux density to adequately constrain a Gaussian fit; the flux density 
reported assumes the source is unresolved.}
\tablenotetext{f}{Low elevation observations ($\sim$13$\arcdeg$)
had poor phase calibration, resulting in degraded astrometry; self-calibration
was possible to retain precise photometry.}
\tablenotetext{g}{Phase stability was too poor to successfully phase reference.
However, the flux for WR~112 was estimated by vector-averaging the bispectrum.  See
\S\ref{section:observations} for more details.}
\end{center}
\end{table}

New positions and the associated uncertainties
of the Wolf-Rayet stars were estimated by taking the mean and standard
deviation of the right
ascension and declination values found in Table\,\ref{table:wrjournal},
excluding data with unusually large errors (due to low elevation observations 
or calibrators with a poorly known positions).  
Our reported uncertainties are conservative, being 
based on the dispersion of the measurements rather than the uncertainty in the
mean, 
since we lack sufficient 
independent observations to properly average
over observing conditions.  
Table\,\ref{table:wr_positions} contains these results along
with the previously determined positions.  The positions for WR~104 and
WR~98a are significantly improved by this work, since the previous
coordinates were based on optical and infrared observations, suffering
from $\simge$1\arcsec errors.

\begin{table}[thb]
\begin{center}
\scriptsize
\caption{Radio positions of Wolf-Rayet stars \label{table:wr_positions} }
\begin{tabular}{lcc|cc}
\tableline\tableline
 & \multicolumn{2}{c}{New Radio Position\tablenotemark{a}\quad (J2000)} & 
   \multicolumn{2}{c}{Previous Position\tablenotemark{b}\quad (J2000)} \\
Source   &    RA  & Dec    & RA & Dec \\
\tableline
WR 104	 & 18 02 04.123 $\pm$ 0.009 & -23 37 42.24 $\pm$ 0.11 
         & 18 02 04.07 & -23 37 41.2  \\
WR 98a   & 17 41 13.051 $\pm$ 0.006 & -30 32 30.34 $\pm$ 0.07 
         & 17 41 12.9  & -30 32 29  \\
WR 112\tablenotemark{c}
         & 18 16 33.489 $\pm$ 0.003 & -18 58 42.47 $\pm$ 0.19 
         & 18 16 33.49 & -18 58 42.5 \\  
\tableline
\end{tabular}
\tablenotetext{a}{This work.}
\tablenotetext{b}{From the VIIth Catalogue of Wolf-Rayet Stars by
\citet{vdHucht2001}.  WR 104 and WR 98a positions were from optical
observations while the position of WR 112 was derived from radio work.}
\end{center}
\end{table}

\section{Modeling}
\label{section:modeling}
\subsection{Methodology}

The radio data for each source were fit with a composite spectral
model, consisting of a thermal wind source ($F_\nu^T$) with non-thermal
emission ($F_\nu^{NT}$) absorbed by a variable amount of
free-free opacity ($\tau_\nu$) from the overlying ionized wind.
We note that $F_\nu^T$ is directly related to the total optical depth of the
wind, but that in this model $\tau_\nu$ represents only the 
optical depth along the line-of-sight to the non-thermal emission 
region $F_\nu^{NT}$, which
clearly must be less than or equal to the total opacity of the wind.
This basic model has been used recently by several workers
\citep{chapman99,skinner99}, and can be expressed in the following
form (following Chapman):
\begin{equation}
F_\nu = F^T_{\nu,\mbox{\rm 4.8\,GHz}} 
\left( \frac{\nu}{\mbox{\rm 4.8\,GHz}}\right)^{\alpha^T} +
F^{NT}_{\nu,\mbox{\rm 4.8\,GHz}} \left( \frac{\nu}{\mbox{\rm 4.8\,GHz}}\right)^{\alpha^{NT}}
e^{-\tau_\nu} 
\end{equation}
where $F_{\nu,\mbox{\rm 4.8\,GHz}}$ refers to the flux density at
4.8\,GHz.

For low signal-to-noise ratio (SNR) 
data of the weak sources (WR~104 and WR~98a), the free
parameters of this model cannot all be independently
well-constrained.  Fortunately, the spectral indices for thermal and
non-thermal emission have been measured for other Wolf-Rayet systems.
We fix the spectral index of thermal emission ($\alpha^T$) to be 0.65,
which is representative of the range observed and predicted by theory
\citep{lr91,williams90,pf75,wb75,olnon75}.  We also fix the spectral index of
the non-thermal ($\alpha^{NT}$) emission to -0.7, based on the range
of values (-0.5 to -1.0) observed around other colliding wind sources
\citep{skinner99,chapman99,dougherty96,setia2000,setia2001}.  
The value of -0.5 can be
physically motivated by considering Fermi acceleration of energetic
electrons in the colliding wind shocks \citep{bell78} and subsequent
synchrotron emission \citep[for more detailed discussion,
see][]{eu93}.  Lastly, we fix the free-free opacity law to be
$\tau_\nu =\tau_{\mbox{\rm 4.8\,GHz}}\left(\frac{\nu}{\rm
4.8\,GHz}\right)^{-2.1}$, which is appropriate for an electron
temperature of $\sim$10$^4$\,K \citep[see Eq. 3-57 in][]{spitzer78}.

Table\,\ref{table:modelfits} contains the modeling results for our target stars,
including 5\% amplitude calibration errors (only important for WR~112).
Because of the low SNR for WR~104 and WR~98a, the best-fitting parameters are
not very meaningful, so
we also report the full range of parameters values consistent
with the data (i.e., the parameter range satisfying the condition that the 
reduced $\chi^2<1$).

\begin{table}[t]
\begin{center}
\scriptsize
\caption{Results from two-component model fitting\label{table:modelfits}}
\begin{tabular}{l|cc|cc|cc|cc}
\tableline\tableline
\multicolumn{1}{c|}{Model} & \multicolumn{2}{c}{WR 104} &\multicolumn{2}{c}{WR 98a} & \multicolumn{4}{c}{WR 112} \\
\multicolumn{1}{c|}{Parameters} && &&  & \multicolumn{2}{c}{Sep 1999} & \multicolumn{2}{c}{Feb 2000} \\
\tableline
$F^T_{\nu,\mbox{\rm 4.8\,GHz}}$ (mJy) & 0.072 & [0 - 0.36]\tablenotemark{a} & 0.078 & [0 - 0.24] 
 & 1.14 & [0.97 - 1.32] & 1.18  & [-]\tablenotemark{b}  \\
$\alpha^T$ & 0.65& [Fixed] & 0.65& [Fixed] & 0.65& [Fixed] & 0.65& [Fixed] \\
$F^{NT}_{\nu,\mbox{\rm 4.8\,GHz}} $ (mJy) & 2.42& [0.78 - 3.70] & 1.14 & [0.33 - 1.96] & 3.47 & [3.01 - 3.94] & 3.16 &[-] \\
$\alpha^{NT}$ & -0.70& [Fixed] & -0.70 &[Fixed] &-0.70 &[Fixed] &-0.70& [Fixed] \\
$\tau_{\mbox{\rm 4.8\,GHz}}$ & 2.47&[0.18 - 3.85] & 1.40 & [0.38 - 2.43] 
& 0.10 & [0.088 - 0.11] & 0.11 & [-] \\
Reduced~$\chi^2$ & 0.024& & 0.040 & &0.63 & & 1.75 &  \\
\tableline
\end{tabular}
\tablenotetext{a}{The best fitting parameters appear in the table, while 
the range of acceptable values follow in brackets (as judged by a
final reduced $\chi^2 < 1$).}
\tablenotetext{b}{The best-fitting model had a reduced $\chi^2 > 1$.}
\end{center}
\end{table}

\subsection{WR 104}
\begin{figure}[t]
\begin{center}
%\epsscale{.5}
\includegraphics[angle=90,width=6in]{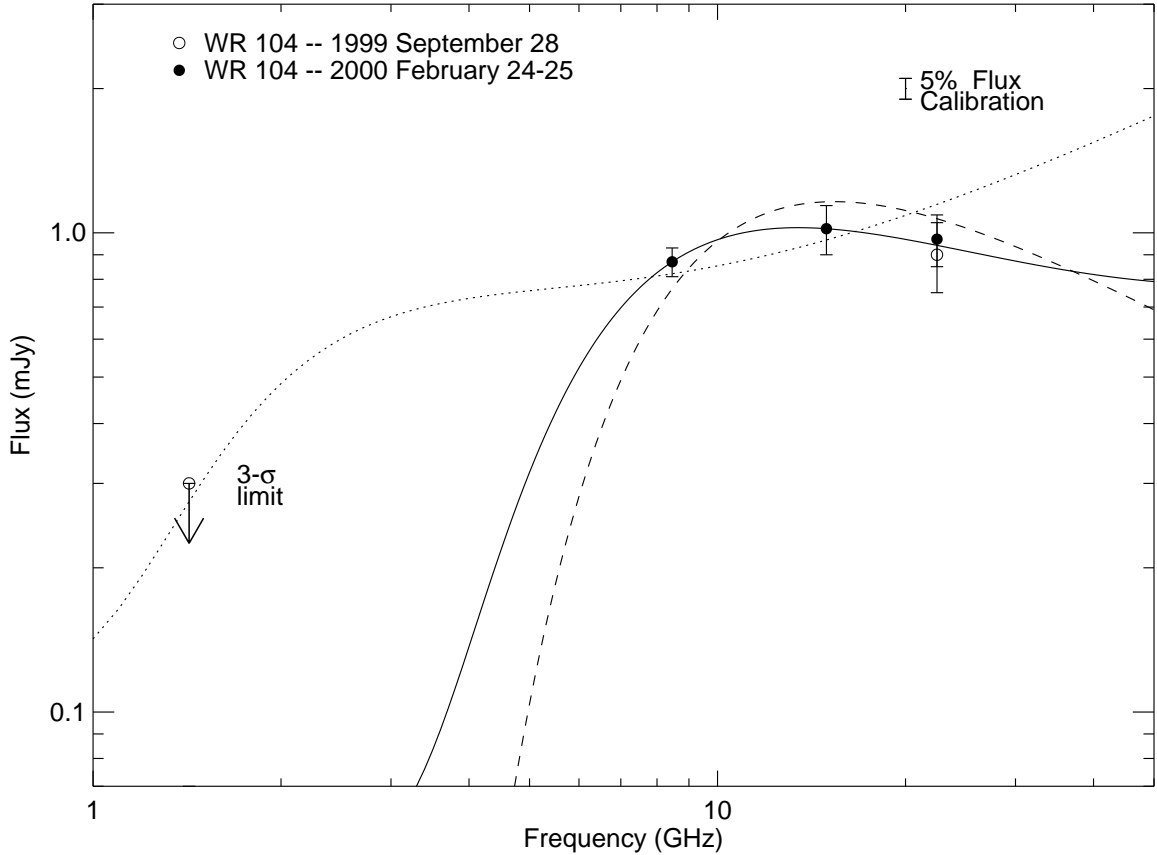}
%      \plotfiddle{Figure/fig:astrometry.eps,rot=90}^M
%\plotfiddle{Figures/fig:astrometry.eps}{12cm}{1}{1}{0}{0}{-90}^M
\caption{Broadband spectra of WR 104 with model fits.  The
best fitting model (reduced $\chi^2$=0.024) 
appears as a solid line while 
two other models also consistent with the data 
(reduced $\chi^2$=1) appear as a dotted line
(tuned for maximum thermal emission, minimum non-thermal emission, 
minimum free-free absorption) 
and a dashed line (tuned for zero thermal emission, maximum non-thermal emission, 
maximum free-free absorption).  See Table\,\ref{table:modelfits} for
the values of the model parameters.
\label{fig:wr104}}
\end{center}
\end{figure}

Figure~\ref{fig:wr104} contains the modeling results for WR~104.
A pure thermal model ($\alpha^T=0.65$) is not a good fit 
(reduced $\chi^2=3.5$); there must be some non-thermal emission.
This result is robust to the potential miscalibration at high frequencies
using Sgr~A$\ast$ (discussed in \S\ref{section:observations}), since
potential overestimate of the calibrator flux density 
would cause our estimate of the
spectral index to be too high (i.e., more similar to thermal emission).

The maximum possible thermal emission component that can be supported
by these new data is 0.36\,mJy at 4.8~GHz.  Interestingly, the data
are consistent with {\em no} detectable thermal emission for models with large
free-free optical depth.  Unfortunately, the optical depth is not very
well-constrained for WR~104 because the spectrum lacks reliable low-frequency
detections.
Quantitatively, the optical depth to the non-thermal
emitting region at 4.8~GHz can range
between 0.18 and 3.85 (best fitting value of 2.47) and still be
consistent with the data. The resulting spectra from these scenarios,
along with the best fitting spectrum, are included in
Figure~\ref{fig:wr104}.

We note that  WR~104 was searched for
radio emission by \citet{leitherer97} and \citet{chapman99} 
who reported 3-$\sigma$ upper limits of 1.59, 0.99, 2.01, \& 0.39\,mJy at
1.38, 2.38, 4.8, and 8.64\,GHz respectively.
Except for the 8.64\,GHz observations, these
results are consistent with the weak detections reported here.

\subsection{WR 98a}
\begin{figure}[t]
\begin{center}
%\epsscale{.5}
\includegraphics[angle=90,width=6in]{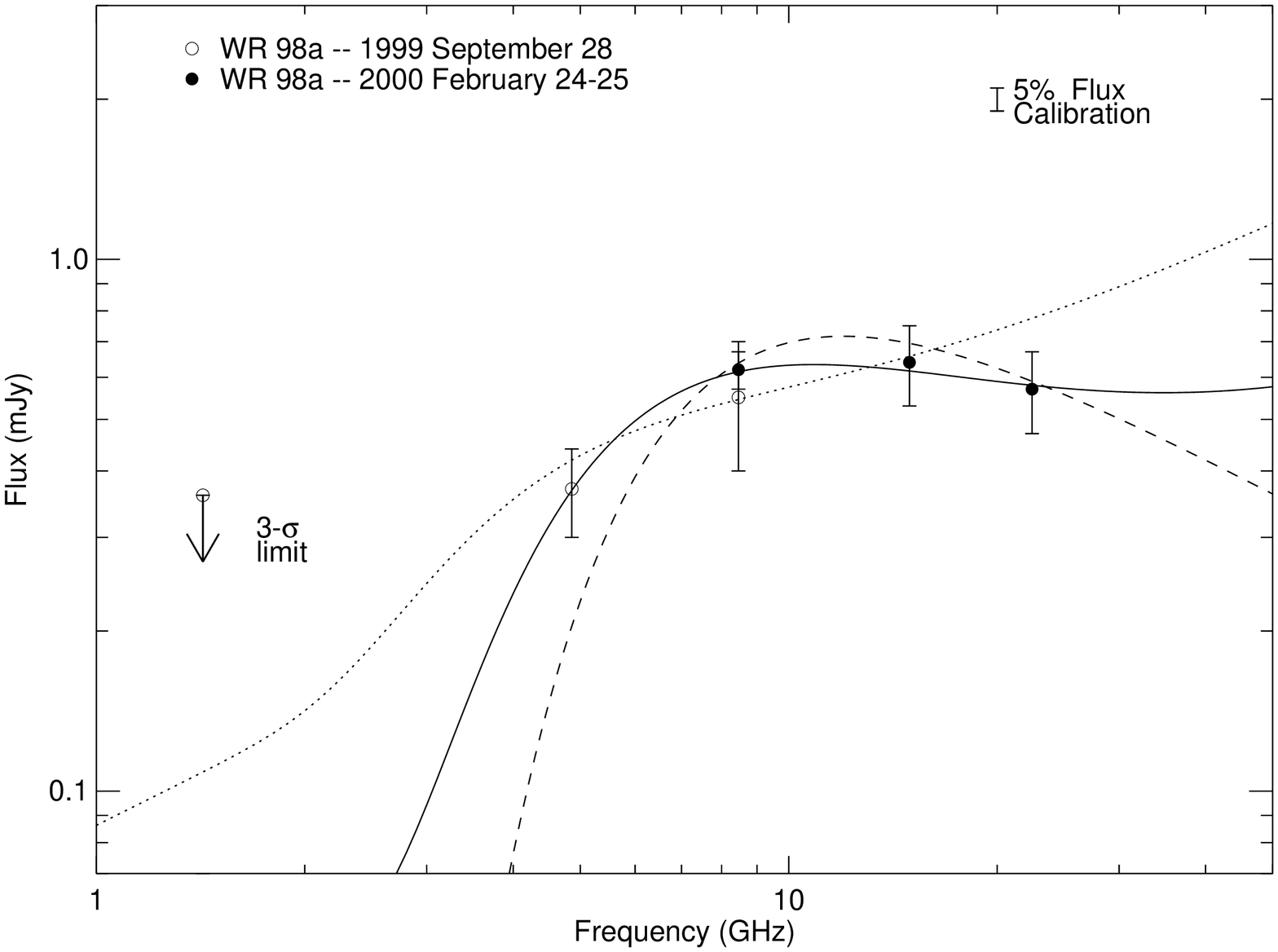}
%      \plotfiddle{Figure/fig:astrometry.eps,rot=90}^M
%\plotfiddle{Figures/fig:astrometry.eps}{12cm}{1}{1}{0}{0}{-90}^M
\caption{Broadband spectra of WR 98a with model fits.  The
best fitting model (reduced $\chi^2$=0.040) 
appears as a solid line while 
two other models also consistent with the data 
(reduced $\chi^2$=1) appear as a dotted line
(tune for maximum thermal emission, minimum non-thermal emission)
and a dashed line 
(adjusted for zero thermal emission, maximum non-thermal emission, 
maximum free-free absorption).  See Table\,\ref{table:modelfits} for
the values of the model parameters.
\label{fig:wr98a}}
\end{center}
\end{figure}

Figure~\ref{fig:wr98a} contains the modeling results for WR~98a.
WR~98a and WR~104 have qualitatively similar spectra,
although a weak 4.86\,GHz detection of WR~98a allows composite spectra models
to be better constrained.

As for WR~104, a pure thermal model ($\alpha^T=0.65$) is not
consistent with these data (reduced $\chi^2$=2.8), and the maximum
thermal emission at 4.8~GHz supported by the data is 0.24\,mJy.  The
data are consistent with {\em no} detectable 
thermal emission for large free-free
optical depths to the non-thermal emission region.  These data
constrain the optical depth to the non-thermal emitting region (at
4.8\,GHz) to lie between 0.38 and 2.43 (best fit at 1.40), based
largely on the 4.86\,GHz detection.

\subsection{WR 112}
\label{SS:wr112}

WR~112 was definitively detected by both \citet{leitherer97} and
\citet{chapman99}, and marked variability was observed between 1995
and 1997, probably caused by varying non-thermal radio emission.
Our observations
support this interpretation and reveal that WR~112 was in a
radio-bright state between 1999 September and 2000 February.  However
at these recent epochs, we do not see evidence for the sharp drop
(more than factor of 3) in flux density between 2.38 \& 1.38\,GHz reported
by \citet{chapman99}.

\begin{figure}[t]
\begin{center}
%\epsscale{.5}
\includegraphics[angle=90,width=6in]{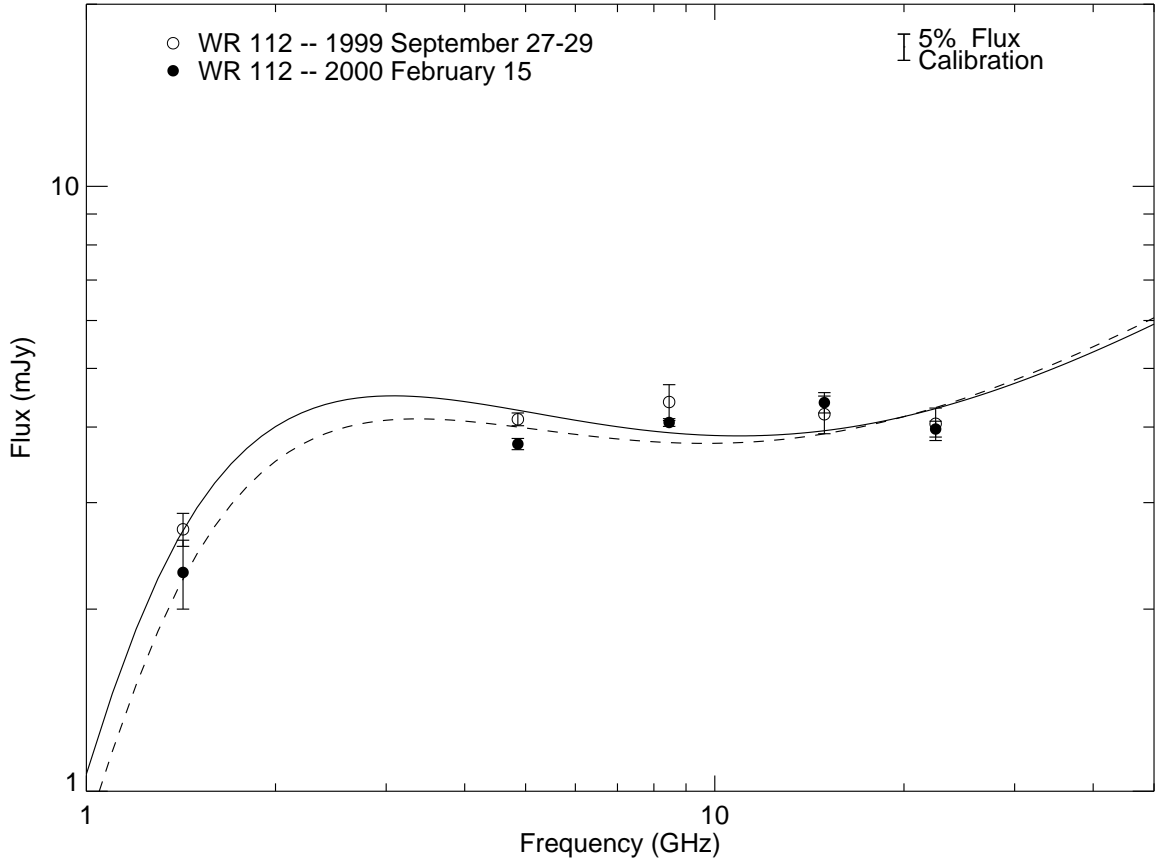}
%      \plotfiddle{Figure/fig:astrometry.eps,rot=90}^M
%\plotfiddle{Figures/fig:astrometry.eps}{12cm}{1}{1}{0}{0}{-90}^M
\caption{Broadband spectra of WR 112 with model fits for two epochs.
The
best-fitting model for September 1999 (reduced $\chi^2$=0.63) 
appears as a solid line while 
the best-fitting model for February 2000 (reduced $\chi^2$=1.75)
appears as a dashed line.
See Table\,\ref{table:modelfits} for
the values of the model parameters.
\label{fig:wr112}}
\end{center}
\end{figure}

Figure~\ref{fig:wr112} contains our modeling results for WR~112.  At
high frequencies, WR~112 shows a flat spectrum similar to WR~104 and
WR~98a.  However, there appears to be
significantly less wind optical depth to the non-thermal emitting
region ($\tau_{\rm{4.8\,GHz}}=0.10$) here than for WR~104 and WR~98a.  
This supports the hypothesis
that WR~112 is a longer period system than WR~104 and WR~98a, already
suspected based on IR and radio variability considerations.  There is
some evidence for a decrease in the low-frequency flux density between 1999
September and 2000 February, corresponding to a $\sim$10\% decrease in
the non-thermal emission component.
We note that subsequent and on-going monitoring of WR~112 by our group
confirms the trend of decreasing non-thermal emission with time, and
full analysis of these variations will be considered in a future
paper.

The high SNR data from 2000 February cannot be fitted by our simple
model within the expected uncertainties (minimum reduced
$\chi^2$=1.75), where the 5\% amplitude calibration error generally
dominates over the statistical measurement error.  
If we allow the spectral indices of the non-thermal and
thermal components to deviate from their fixed values, the fit improves 
but is still not satisfactory.  Specifically, if we insist that
$\alpha^T$ lie between 0.6 and 0.7 and that $\alpha^{NT}$ stay
between -0.5 and -1.0, the best fit model has a reduced $\chi^2$=1.20
(for $\alpha^T$=.6 and $\alpha^{NT}$=-0.5).

\citet{skinner99} \citep[see also][]{setia2001} 
fit the radio spectrum of long-period,
colliding-wind binary WR~147 over a similar frequency range
and found that an absorbed monoenergetic
synchrotron spectrum fit the data better than the absorbed power-law
model considered here.  This better fit was largely due to a steep
high-frequency cutoff observed in the non-thermal spectrum around
22~GHz which is more naturally accommodated by the synchrotron spectrum.  
A high-frequency cutoff is 
intimated  in our WR~112 data but is 
uncertain due to difficulty in estimating the strength of the thermal
emission.  \citet{skinner99} were able to more reliably estimate the
thermal emission component  using 43~GHz observations and the fact
that the thermal and non-thermal components were spatially resolved 
from each other at both 15~GHz and 22~GHz.

\section{Discussion}

\begin{table}[thb]
\begin{center}
\scriptsize
\caption{Derived parameters based on radio spectra\label{table:derived} }
\begin{tabular}{l|cccccl|cc}
\tableline\tableline
Source	& Distance & $v_\infty$ & $a$\tablenotemark{a}    & $F^T_{\mbox{\rm 4.8\,GHz}}$  &
  Max & Refs &  \.{M} & R$^{\rm{NT}}$\tablenotemark{b} \\
	& (kpc)    & (km s$^{-1}$) & (AU) & (mJy)   &   $\tau_{\mbox{\rm 4.8\,GHz}}$
  & & (10$^{-5} \msun$ yr$^{-1}$) & (AU) \\
\tableline
WR 104 (WC9)   & 2.3 & 1220 & 2.4 & 0.072 ($<0.36$)\tablenotemark{c} & 3.9 & 1,4
& 0.8 ($<2.8$) & $>$5 \\
WR 98a (WC8-9) & 1.9 & 900  & 4.1 & 0.078 ($<0.24$) & 2.4  & 2,4
& 0.5 ($<1.1)$ & $>$4 \\
WR 112 (WC9)   & 1.3 & 1100 & 6.1\tablenotemark{d} & 1.16$\pm$0.17   &0.11 & 3,4
& 2.5$\pm$0.3\tablenotemark{e} & $\sim$46\tablenotemark{f}\\
\tableline
\end{tabular}
\tablenotetext{a}{Binary separation based on the orbital period and 
the assumption that both components are 15\,\msun.}
\tablenotetext{b}{Transverse distance from shock stagnation point 
to location 
corresponding to wind opacity
$\tau_{\mbox{\rm 4.8\,GHz}}$; estimate of the intrinsic size of 
the non-thermal
emission (see discussion in \S\ref{section:nt})}
\tablenotetext{c}{Quantities in parentheses represent upper limits based on
fits to radio spectra.}
\tablenotetext{d}{Period of WR~112 binary is unknown, but estimated to be
$\sim$1000\,days based on unpublished radio and IR data.}
\tablenotetext{e}{This error bar only reflects errors in determining
$F^T_{\mbox{\rm 4.8\,GHz}}$, and not the much larger uncertainties in the
distance and wind velocity.}
\tablenotetext{f}{This assumes line-of-sight does not traverse the
OB-star wind, but only the WR wind.}
\tablerefs{1. \citet{tuthill99}. 2. \citet{monnier99}. 
3. \citet{nugis98}. 4. This work. }
\end{center}
\end{table}

\subsection{Thermal emission and mass-loss rates}
The thermal emission from Wolf-Rayet stars
is understood to arise from
free-free emission in an expanding, ionized wind \citep{wb75,pf75,olnon75}.
Following \citet{wb75}, the observed flux density can be related to
stellar and wind parameters as follows:
\begin{equation}
F^T_\nu = 2.32\times 10^{4} 
  \left( \frac{\dot{M}Z}{v_\infty \mu}\right)^{\frac{4}{3}}
  \left( \frac{\gamma g_\nu \nu}{d^3} \right) ^{\frac{2}{3}}
\label{eq:mdot}
\end{equation}
where $F^T_\nu$ is the observed flux density 
in mJy, $\dot{M}$ is the mass-loss rate
in \msun\,yr$^{-1}$, $Z$ is the mean ionic charge, $v_\infty$ is the wind
velocity in km\,s$^{-1}$, $\mu$ is the mean molecular weight, $\gamma$
is the mean number of electrons per ion, $g_\nu$ is the free-free Gaunt 
factor, $\nu$ is the observing frequency in Hz, and $d$ is the distance in kpc.

\citet{leitherer97} discuss the best values to use for late-WC
stars, and we have followed these authors in adopting $Z=1.1$,
$\gamma=1.1$, $\mu=4.7$, and $g_{\rm{4.8~GHz}}=5.03$ in the
calculations that follow.  Table~\ref{table:derived} contains the
estimated distances and wind velocities of our targets derived from the
literature, and fluxes from thermal emission
at 4.8~GHz (from Table\,\ref{table:modelfits}) based on the 
spectral decompositions
of \S\ref{section:modeling}.
With these data, equation (\ref{eq:mdot}) can be inverted to solve for
the mass-loss rate, and these results are also contained in
Table~\ref{table:derived}, corresponding to both the best-fitting
values for the thermal emission component and the maximum allowed
value.  Also included in this table is an estimate of the major axis of the
binary system based on the period and assumption of 15\,\msun components.

The estimated mass-loss rates span 0.5 to 2.8 $\times
10^{-5}~\msun$\,yr$^{-1}$, similar to results from previous studies
of late-WC WR stars.
\citep[e.g.,][]{nugis98,leitherer97,abbott86}.  Because of the presence
of non-thermal emission in all our sources, our determinations of
mass-loss rate are very uncertain and we have neglected secondary
effects such as clumping corrections \citep{moffat94,morris2000}.  
Interestingly, WR~112 seems to show a markedly higher (2.5$\times$)
mass-loss rate than observed in 1995 \citep{leitherer97} -- that is,
the high frequency radio
emission (most likely dominated by thermal emission)
was significantly higher in the more recent epochs.
Although the
orbital properties of the WR~112 binary are unknown, it is not
expected that the thermal emission, which is dominated by free-free
emission in the Wolf-Rayet wind, should be a strong function of the
binary separation or other parameter.
Higher
frequency observations at multiple epochs 
can better isolate the thermal emission from
non-thermal emission, critical for accurately estimating the mass-loss
rates for individual colliding wind sources.

\subsection{Non-thermal emission}
	
As previously discussed in \S\ref{SS:wr112}, the physical mechanism
producing the non-thermal radio emission is not well known, and
predictions from many models can be compared to the observed spectra
\citep[e.g.,][]{skinner99,setia2000,setia2001}.  
Despite these uncertainties, one component
that all non-thermal emission models 
contain is free-free absorption by the overlying ionized 
wind.
The free-free optical depth to the non-thermal emission
implied by power-law models (as considered
here) is generally larger than that obtained by fitting to alternative
model spectra whose non-thermal source contain an intrinsic low-frequency
turnover.
Hence we can
interpret the optical depth parameters derived in the last section to
be reasonable upper limits.  We can therefore estimate a lower limit on wind
depth where the emission arises, assuming only that the intrinsic
non-thermal spectrum has low-frequency behavior flatter than the -0.7
spectral index used here for fitting (this includes the absorbed
monoenergetic synchrotron spectrum preferred by Skinner\etal 1999).

\begin{figure}[t]
\begin{center}
%\epsscale{.5}
\includegraphics[angle=0,width=6in]{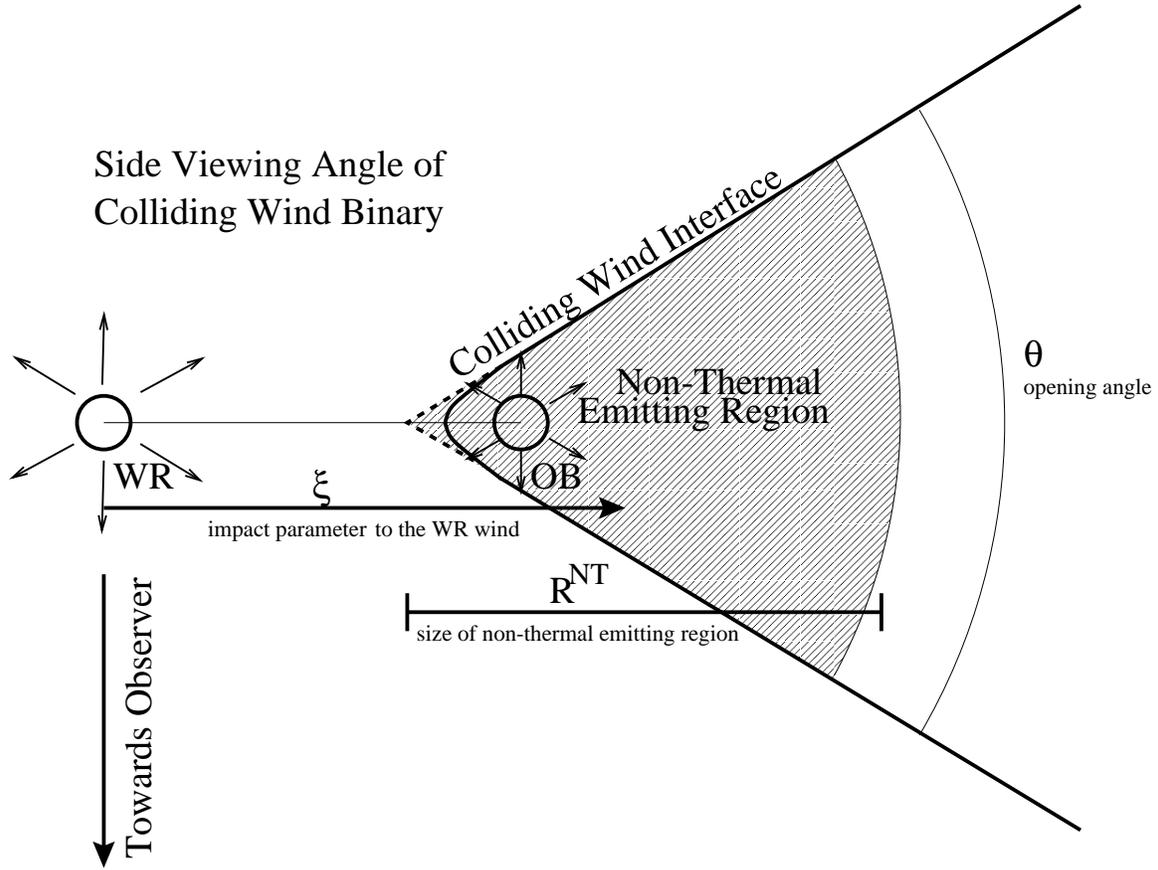}
%      \plotfiddle{Figure/fig:astrometry.eps,rot=90}^M
%\plotfiddle{Figures/fig:astrometry.eps}{12cm}{1}{1}{0}{0}{-90}^M
\caption{Schematic of colliding wind geometry for WR+OB binaries
(not to scale).
The intrinsic size of the non-thermal emitting region can be
estimated from the observed lower limits on the wind optical depth.
See Table\,\ref{table:derived} for
the values of the model parameters. 
\label{fig:schematic}}
\end{center}
\end{figure}

In order to calculate the optical depth of the overlying wind for variously
sized non-thermal emitting regions, we must first 
discuss how a binary system alters the wind density around the WR star.
Let us consider a colliding wind system viewed nearly face-on, as is
appropriate for the pinwheel nebulae WR~104 and WR~98a.  The
non-thermal radio emission originates at the interface of the
colliding winds, which has a nearly conical geometry outside the
interaction region (see Figure\,\ref{fig:schematic}) with  
an opening angle, $\theta$, dependent
on the relative WR and O-star wind momenta.  For wind momenta ratio
between 0.01 and 0.1 (expected for these systems) 
the opening angle
$\theta$ lies between 50\arcdeg and 90\arcdeg \citep{eu93}, 
consistent with the geometrical thickness
of the dust plume seen around WR~104.  In order to estimate the
minimum intrinsic size of the non-thermal emission, we must calculate
the impact parameter, $\xi$, which would intersect this cone with
$\tau_{\rm{4.8~GHz}}\sim 1$.  Because the integrated line-of-sight
optical depth of the wind falls off steeply, $\tau\propto\xi^{-3}$
\citep{wb75,pf75}, this estimate is relatively insensitive to wind
parameters.  See Figure\,\ref{fig:schematic} for a sketch of the wind
and orbital geometry.

For a generic late-WC wind with $\dot{M}=1\times10^{-5}\,\msun\,$yr$^{-1}$
and wind speed of 1000\,km\,s$^{-1}$, the $\tau_{\rm{4.8~GHz}}\sim 1$ 
condition is reached for $\xi\sim$ 18, 15, 13 AU for opening angles 
$\theta\sim$ 0\arcdeg, 60\arcdeg, 
90\arcdeg respectively.  For larger opening angles, the non-thermal emission
extends more toward the observer and is visible at smaller
impact parameters.  Hence to be observed at all, the intrinsic non-thermal 
emission region must extend along the interface cone to this 
distance $\xi$. This size is significantly larger than the
estimated non-thermal emission region size estimated by \citet{eu93}, which
was largely based on what scale the wind collision can efficiently put
energy into the
shocks (1 to 3 AU for our target stars).

It has been pointed out by previous investigators that the estimated
optical depth to the shock collision zone is so large that all
non-thermal radio emission should be completely obscured for periods
$\simle$2\,yr \citep{dw2000,chapman99,wb95,williams90,williams94}.  Most
recently, \citet{dw2000} discuss a number of explanations for why
nonthermal emission is visible for WR~11 ($\gamma^2$~Vel, period
78.53~days).  At radio wavelengths,
one can see deeper into the wind
if it is clumpy \citep[e.g,][]{nugis98} or non-spherical
\citep{williams97}.  In addition for nearly edge-on systems, 
lines-of-sight can pass through the relatively less-dense O-star wind
during some parts of the orbit, causing the observed 
radio spectrum to vary.

Since the WR~104 and WR~98a binaries are viewed within $\sim$35\arcdeg~~from
face-on \citep{monnier99}, we can eliminate possible explanations
invoking novel observing geometries, such as viewing the collision
zone through the O-star wind.  Further, the wind would need to be very
clumpy and/or non-spherical to allow emission at $\xi\sim$1-3\,AU to
be visible, considering that $\tau_\nu \propto \xi^{-3}$; the 4.8\,GHz
optical depth in the smooth wind to the collision zone is $\sim$50 
for typical late-WC stars.
Such radical departures from smooth flow would likely be seen as
significant changes to the thermal emission, which arises from the
same ionized wind material.

Here we develop another possibility, that the non-thermal emission
region may be intrinsically larger than expected \citep[also suggested
in part by][]{dw2000}.  The plasma could continue to radiate beyond the
initial collision region for at least the synchrotron cooling
timescale.  In order for the plasma to travel out of the collision
region and reach the $\tau_{\rm{4.8~GHz}}\sim 1$ surface, this time
would need to be $\simge$10\,days, a reasonable time span considering
the analysis of \citet[see \S3]{eu93}, assuming magnetic fields of a
few Gauss.  The size and strength of the radio emission ``tail'' also
depends greatly on the magnetic field geometry; for uniform magnetic
fields the energetic electrons can escape to greater distances from
the collision region than for non-uniform magnetic fields, and could
produce observable radio emission (Usov 2001, private communication);
thus these observations might be revealing important new information
regarding the magnetic fields of WR+OB binaries.  Interestingly,
\citet{corcoran96} found evidence that the X-ray emitting region of
the colliding wind system V444~Cyg (WR~139), presumed to trace hot
shocked gas, was also unexpectedly large (10$\times$ the binary
separation).  Lastly, it might be possible that even compact
non-thermal radio emission may appear larger due to electron scattering in the
dense ionized wind.

We have calculated the minimum impact parameter $\xi$ for each of our
target stars based on the (best-fit) estimated mass-loss rates, the
orbital and wind parameters, the upper-limit of $\tau_{\rm{4.8\,GHz}}$
derived from our spectral fits, and a cone angle of 60\,\arcdeg. From
this minimum $\xi$, we can estimate the minimum extent of the
non-thermal emission, R$^{\rm{NT}}$, by subtracting the binary
separation from the minimum $\xi$ (since the dense WR wind is centered on the
WR star while the collision region occurs close to the OB-star).
These values have been placed in
Table\,\ref{table:derived} and should be taken as lower limits to the
intrinsic size of the non-thermal emission region; the non-thermal
emission region could be much larger since we are adopting
conservative upper-limits for the optical depth.

This model can be tested by high-resolution
imaging of the non-thermal radiation, but VLBA observations of WR~104
and WR~98a are currently not practical due to their weak flux.  While WR~112 is
bright enough to detect (in its radio-bright state), the closest phase
reference calibrator is $\sim$5\arcdeg~~away.  Initial attempts to
image the non-thermal radio emission (by our group) have failed either
because the source was over-resolved or due to
spatial incoherence effects attributed to low elevation and the
large angular distance between the WR~112 and the phase reference.
VLBA observations of closer colliding wind sources, such as WR~140
near periastron or WR~11, should be able to resolve these structures
and determine precisely the physical geometry of the non-thermal
emitting regions.

\label{section:nt}

\subsection{Effects of dust on the radio emission}
\begin{figure}[t]
\begin{center}
%\epsscale{.5}
\includegraphics[angle=0,width=6in]{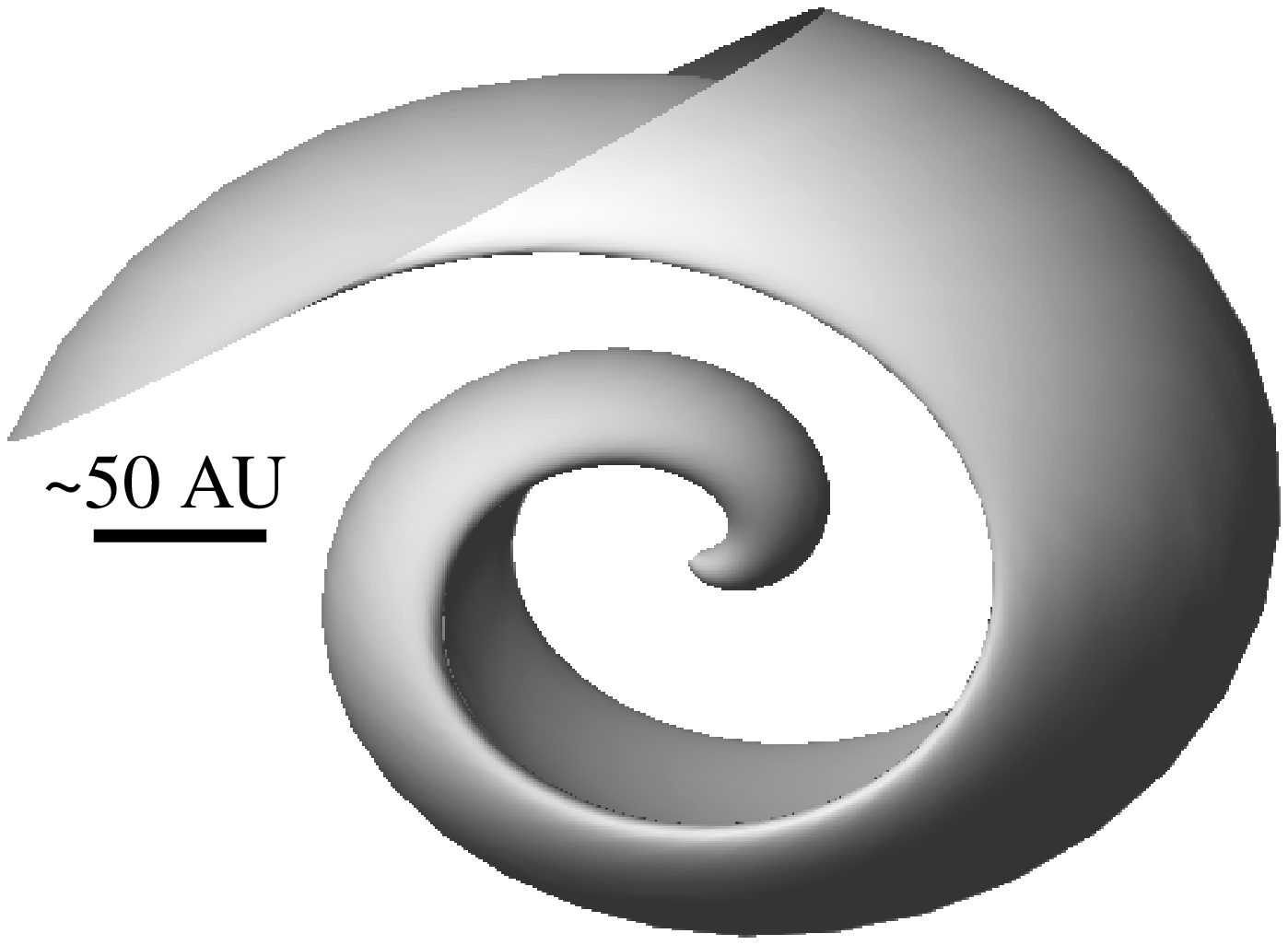}
%      \plotfiddle{Figure/fig:astrometry.eps,rot=90}^M
%\plotfiddle{Figures/fig:astrometry.eps}{12cm}{1}{1}{0}{0}{-90}^M
\caption{Three-dimensional representation of the colliding wind
interface for WR~104. In this picture, the two stars are located in
the very center separated by a few AU horizontally.  As the stars
orbit (clockwise), the shock cone wraps around into a thin, quasi-spherical
sheet.  The termination of this surface is defined for material which
left the WR star at the same time.  If the dust exists in a thin sheet
at this interface, the optical depth could be sufficient to block
stellar ultra-violet photons from ionizing the wind and thus reduce the
observed free-free emission.
\label{fig:spiralmesh} }
\end{center}
\end{figure}

If the dust formed at the colliding wind interface is optically thick to
UV-photons, a significant fraction of the circumstellar gas will be
shielded from the ionizing flux of the central sources.  
Figure\,\ref{fig:spiralmesh} shows an idealized model of 
colliding wind interface geometry for WR~104
(for opening angle $\theta=90$\arcdeg). 
The radius of the WR 104 radio photosphere is $\sim$15\,AU at 4.8\,GHz
\citep[$\propto\nu^{-0.66}$,][]{wb75}, which is similar to the length
scale of the spiral wind interface in Figure\,\ref{fig:spiralmesh}.  
Hence, dust in the interface region can absorb the UV flux and
shadow a significant volume of the wind (especially in the mid-plane).  
Initial calculations
indicate that the thermal emission will be smaller by $\sim$30\% for
the same mass-loss rate, and the spectral slope might be slightly
larger than the canonical $\alpha^{\rm{T}}\sim0.65$.  
We note that these effects may even be important for spherical distributions
of grains, since the inner radius of the dust shell in such models is also
$\sim$25~AU \citep{zubko98} and could change the ionization fraction in the
outer envelope from standardly-used values.

The brightness of the non-thermal radio emission could also be
affected.  For binary systems viewed nearly edge-on, the dust shadow
would also significantly reduce the line-of-sight opacity to the
collision zone (since the gas will not be ionized in the shadow) at
certain orbital phases than currently expected.  More work is needed
to quantitatively understand these effects, and will need to include a
detailed treatment of the dust opacity as well as recombination time
scale for ``shadowed''-gas.

Such effects could be detected with high resolution images of the thermal 
emission region, since the shadowing would cause the emission to
be very asymmetrical.  By observing at mm-wave wavelengths, it would
be possible to probe deeper
layers on the wind, revealing the structure of the
wind-wind interface on $\sim$10\,AU scales (e.g., using MERLIN, ALMA).  
In addition, signs of shadowing might also be seen on larger scales
in nebular emission lines or as linear polarization at visible wavelengths.

\subsection{Time-variability} 
Unfortunately, all our observations
of individual sources at different frequencies
were not
taken at the same epoch, which is potentially troubling since WR~104
and WR~98a are known to have binary periods of 243.5 and 565 days
respectively \citep{monnier99}.  Fortunately, there is some wavelength
overlap between the epochs for both sources and these flux
measurements are consistent to within errors.  This is not surprising
for WR~104 since analysis of the spiral morphology indicates a nearly
face-on viewing angle \citep[11$\pm$7\arcdeg ,][]{mythesis}.  There is
some IR flux variation observed for WR~98a \citep{williams95}, either due
to optical depth effects (viewing angle is $\sim$35~degrees for WR
98a) or orbital eccentricity, but we do not detect a similar radio
variation in our limited dataset.  In both cases, the colliding 
wind interface should not cross our line-of-sight and
hence we do not
expect large variations of the kind seen in other systems, such as WR~140
\citep{wb95}.

WR~112 showed evidence for slight variation between 1999 September and
2000 February at low frequencies.  We are monitoring this source with
the VLA in order to understand the large variation originally observed
by \citet{leitherer97} and \citet{chapman99}.  This variation can be
either due to changes in line-of-sight opacity or intrinsic radio
emission, caused by either a large inclination angle or binary
separation (or both).  
Coordinated infrared imaging and broadband
radio spectra will go a long way towards supporting the creation of a
self-consistent model, and clarifying the likely cause(s) of the
observed variability.

\section{Conclusions}
We detected non-thermal radio emission 
from both known pinwheel nebulae (WR~104 and WR~98a) and the
suspected binary system WR~112.
Pure thermal emission
from a spherically-symmetric ionized wind
was not consistent with the spectra from any of the sources, and
some non-thermal emission was always required to explain the observations.
Simple composite spectral models were fit to
the broadband measurements, allowing emission strengths of the thermal
and non-thermal components to be estimated along with the
line-of-sight opacity towards the non-thermal emission region.  
From the inferred thermal emission strengths and known wind properties,
we estimated the mass-loss rates and found them to be similar to those for
other late-WC WR stars, validating the basic spectral decomposition employed
here.

Using these results, the upper bound line-of-sight opacities to
the non-thermal emission were inverted to estimate lower-limit sizes of
the non-thermal emission regions, which were
many times larger than expected
from current theory \citep{eu93}.  We suggest that synchrotron cooling
times of a few weeks would allow the radio-emitting plasma to travel
into an optically thin portion of the thermal WR wind and to be observed, 
or that
uniform magnetic fields could allow energetic electrons to escape out of the
radio ``photosphere.''

We also discussed implications of dust obscuration on the radio
properties of dusty WRs.  If the dust in the outflow is optically
thick to ultraviolet photons, a significant fraction of the wind will
be shadowed from the major sources of ionization.  This will affect
mass-loss rates determined from high frequency radio observations for
all binary (i.e. non-thermal) systems 
and could cause dramatic differences in the
line-of-sight free-free opacity for binary systems viewed edge-on.

A lack of high frequency (e.g., 43~GHz) observations hampered our analysis
since even 22~GHz observations were significantly contaminated by
non-thermal radio emission, making the decomposition of spectra
reliant on {\em a priori} estimates for the model spectral indices.
For WR~112, high SNR measurements suggest that simple
two-component power-law models commonly used are inadequate, hinting at a high
frequency turnover to the non-thermal radio emission.
We emphasize that further progress in this field will require
multi-wavelength simultaneous observations from 1 to 43 GHz in order
to untangle the contributions from the thermal and non-thermal 
components.  

The hypothesis that colliding winds lie at the heart of all Wolf-Rayet
dust shells has passed another observational test, detection of
non-thermal radio emissions from WR~104, WR~98a, and WR~112.  A
sensitive radio survey ($\sigma\simle$0.1~mJy) between 1 and 43~GHz of
all known dusty Wolf-Rayets would put this hypothesis to a final
challenge.

%% The \notetoeditor{TEXT} command allows the author to communicate
%% information to the copy editor.  This information will appear as a
%% footnote on the printed copy for the manuscript style file.  Nothing will
%% appear on the printed copy if the preprint or
%% preprint2 style files are used
\acknowledgments

The authors would like to recognize useful discussions and comments
from S. Dougherty and V. Usov.  This research has made use of the
SIMBAD database, operated at CDS, Strasbourg, France, and NASA's
Astrophysics Data System Abstract Service.  JDM acknowledges support
from a Center for Astrophysics Fellowship at the Harvard-Smithsonian
Center for Astrophysics.

\bibliographystyle{apj}
\bibliography{apj-jour,Radio_Pinwheel,Thesis}

%\begin{thebibliography}{apj-jour,CIT_6,Thesis}
%\end{thebibliography}

%% Generally speaking, only the figure captions, and not the figures
%% themselves, are included in electronic manuscript submissions.
%% Use \figcaption to format your figure captions. They should begin on a
%% new page.

\clearpage

%% No more than seven \figcaption commands are allowed per page,
%% so if you have more than seven captions, insert a \clearpage

%% There must be a \figcaption command for each legend. Key the text of the
%% legend and the optional \label in curly braces. If you wish, you may
%% include the name of the corresponding figure file in square brackets.
%% The label is for identification purposes only. It will not insert the
%% figures themselves into the document.
%% If you want to include your art in the paper, use \plotone.
%% Refer to the on-line documentation for details.

%\figcaption[sgi9259.eps]{This is the first figure and it uses sgi9259.eps as
%its EPS figure file. \label{fig1}}

%% Tables should be submitted one per page, so put a \clearpage before

%% Two options are available to the author for producing tables:  the
%% deluxetable environment provided by the AASTeX package or the LaTeX
%% table environment.  Use of deluxetable is preferred.
%%

%% Three table samples follow, two marked up in the deluxetable environment,
%% one marked up as a LaTeX table.

%% In this first example, note that the \tabletypesize{}
%% command has been used to reduce the font size of the table.
%% Note also that the \label command needs to be placed 

\clearpage

\end{document}